\documentclass[aps,preprint,prb]{revtex4}
\usepackage{graphicx}

\begin{document}

\title{Effect of Particle Shape and Charge on Bulk Rheology of Nanoparticle Suspensions}
\author{David R. Heine}
\affiliation{Science and Technology Division, Corning Incorporated, SP-TD-01-1, Corning, New York 14831, USA}
\author{Matt K. Petersen}
 \altaffiliation[Present address]{Department of Chemistry and Center for Biophysical Modeling and Simulation, University of Utah, Salt Lake City, Utah 84112-0850}
\author{Gary S. Grest}
\affiliation{Sandia National Laboratories, Albuquerque, New Mexico 87185, USA}
\date{\today}

\begin{abstract}
The rheology of nanoparticle suspensions for nanoparticles of various shapes with equal mass is studied using molecular dynamics simulations.  The equilibrium structure and the response to imposed shear are analyzed for suspensions of spheres, rods, plates, and jacks in an explicit solvent for both charged and uncharged nanoparticles.  For the volume fraction studied, $\phi_{\rm vf}=0.075$, the uncharged systems are all in their isotropic phase and the viscosity is only weakly dependent on shape for spheres, rods, and plate whereas for the jacks the viscosity is an order of magnitude larger than for the other three shapes. The introduction of charge increases the viscosity for all four nanoparticle shapes with the increase being the largest for rods and plates. The presence of a repulsive charge between the particles decreases the amount of stress reduction that can be achieved by particle reorientation.
\end{abstract}
\maketitle

$\dagger$ Present address: Department of Chemistry and Center for Biophysical Modeling and Simulation, University of Utah, Salt Lake City, Utah 84112-0850

\section{Introduction}
\label{sec:intro}

With increasing capabilities to synthesize nanoparticles with a wide variety in compositions, shapes and sizes there has been an increasing interest in how these properties influence the behavior of the bulk material.  Understanding what nanoparticle properties most strongly influence the bulk behavior and to what extent the bulk properties vary can provide a significant advantage in a number of industries that process powders, suspensions, and solutions containing nanoscale particles.  One such property is the shape of the nanoparticle which even in naturally occurring materials can vary drastically from elongated rods to near perfect spheres or amorphous blobs.  To what extent do these various shapes impact the rheology of nanoparticle suspensions?  How much of a difference in nanoparticle shape is required to use tailored nanoparticle shapes as viscosity modifiers?  We address these questions by simulating the rheological behavior of suspensions of nanoparticles with different shapes.

\begin{figure}
 \centering
 \includegraphics[clip=false,width=8 cm]{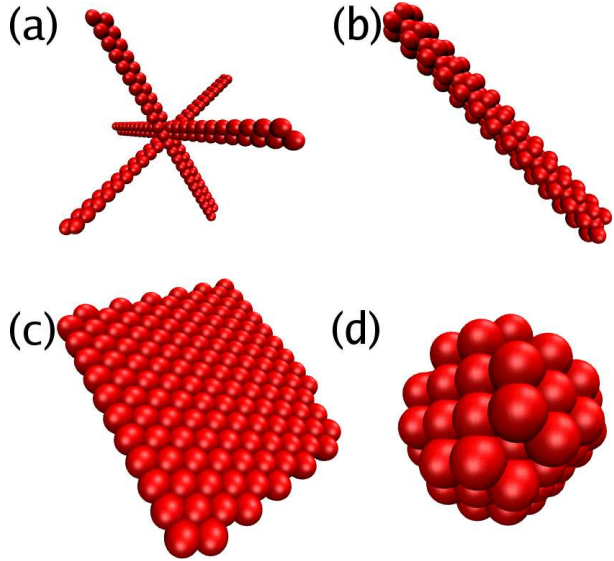}
 \caption{Snapshots of individual particles for (a) jacks consisting of three $25 \sigma$ long and $1.2 \sigma$ wide arms, (b) rods that are $22.5 \sigma$ long and $2 \sigma$ wide, (c) plates with each side $12 \sigma$ long, and (d) spheres with diameters of $5.6 \sigma$.}
 \label{fig:SINGSNAP}
\end{figure}

Multiple approaches have been used to study the rheology of suspensions including Brownian dynamics,\cite{SRR:JCP:96,DRF:JR:00} dissipative particle dynamics,\cite{NSM:JR:05,VP:JCP:05,ESB:PRE:97} and stochastic rotation dynamics (SRD).\cite{MH:PRE:05,MH:PRE:06}  Almost all of these approaches treat the nanoparticles as perfectly spherical particles in either an implicit solvent or an explicit solvent consisting of  spherical particles.  For SRD, the implicit solvent particles are treated as point masses.  Some work has been done using quadrupoles to describe suspensions of platelets representing clay particles instead of spheres\cite{MD:PRE:97,MD:PRL:95,GO:PRE:04} but this approach does not allow for a wide range of particle shapes.  Others have described particles of arbitrary shapes using composites of sub-particles held together as rigid \cite{ZZ:NL:04,MP:PRE:09} or flexible bodies\cite{STK:JPS:07}.  This is the approach we use here to study the effect of nanoparticle shape on suspension rheology for the four nanoparticle shapes shown in Fig. \ref{fig:SINGSNAP}.  In addition, we also consider how the introduction of charge on the nanoparticle affects their suspension rheology in the presence of both low and high salt solutions.  

In Section \ref{sec:details} we define the systems studied including the interaction potentials between nanoparticles and the system properties.  We also describe the techniques used to simulate the equilibrium and sheared systems.  In section \ref{sec:structure} we present our results for the equilibrium structure and diffusion coefficient for nanoparticle suspensions for the four nanoparticle shapes.  We determine the relationship between the diffusion coefficient and ionic strength for charged spheres and jacks as we approach the no salt limit. Results for the orientational order parameter for rods and plates for different ionic strength are also presented. In section \ref{sec:rheology} we present results for the viscosity of the sheared suspensions.  A brief summary of our results and conclusions are presented in Section \ref{sec:conc}.

\section{Model Details}
\label{sec:details}

Nanoparticles in the shape of rods, plates, jacks and spheres were each modeled as rigid composites of 135 interaction sites each of mass $m$ as shown in Fig. \ref{fig:SINGSNAP}. The uncharged double wide jacks discussed here as well as extended jacks consisting of single row of interaction sites have previously been studied by Petersen et al.\cite{MP:PRE:09}
The background solvent consists of Lennard-Jones particles of mass $m$. 
The interaction potential for both the solvent and nanoparticle sites is the Lennard-Jones (LJ) 6-12 potential,
\begin{equation}
u_{LJ}=\left\{
\begin{array}{ll}
4\varepsilon[(\frac{\sigma}{r})^{12}-(\frac{\sigma}{r})^{6}] & r\leq r_{c} \\ 
0 & r>r_{c}
  \end{array}\right.\label{eq:ljpot} 
\end{equation}
where $\varepsilon$ and $\sigma$ are the LJ units of energy and length and are the same for both the nanoparticle interaction sites and the solvent. The solvent-solvent and solvent-nanoparticle site interactions were cutoff at $3\sigma$ while the nanoparticle-nanoparticle site interactions were cutoff at $2^{1/6}\sigma$.  The shorter cutoff for the nanoparticle-nanoparticle site interaction results in a purely repulsive, short range interaction between nanoparticles, thereby promoting nanoparticle dispersion. 
The equations of motion were integrated with a velocity-Verlet algorithm with a time step $\Delta t = 0.005\tau$, where $\tau=\sigma(m/\epsilon)^{1/2}$. The density of the nanoparticles was nominally the same as that of the background solvent to avoid the consideration of sedimentation.
The rigid-body motion was determined by the forces on the nanoparticle's composite sites.  These forces produced center-of-mass translation and rotation about the nanoparticle center-of-mass. All of the simulations for the uncharged systems were carried out at a nanoparticle volume fraction of $\phi_{\rm vf}=0.075$.  This volume fraction was chosen since it was large enough that the suspension viscosity $\eta$ was measurably larger than that of the solvent viscosity, $\eta_s=1.01 \pm 0.03 m/\tau\sigma$,\cite{PITV:PRE:09} and low enough that at least for the uncharged systems the nanoparticles were in their isotropic phase. Although each shape is composed of the same number of sub-particles, the surface area and hydrodynamic radius varies for each type, thus giving rise to shape-dependent rheological behavior.  

The effect of introducing charge on the nanoparticles was also studied by adding a DLVO interaction between all interactions sites on the nanoparticles,
\begin{equation}
u_{DLVO}=\left\{
\begin{array}{ll}
A(\frac{e^{-\kappa r}}{r}) & r\leq r_{c} \\ 
0 & r>r_{c}
  \end{array}\right.\label{eq:dlvopot} 
\end{equation}
where the prefactor A is set to $1.0 \varepsilon$ and the cutoff is at $r_{c}=10\sigma$.  We compare high screening and low screening to the no charge case for all four systems by setting the inverse Debye length to $\kappa=0.34\sigma^{-1}$ for the high screening case and $\kappa=0.086\sigma^{-1}$ for the low screening case.  These values correspond to Debye lengths of $d/2$ and $2d$, respectively, where $d=5.8\sigma$ is the diameter of the spherical composite particle.  Upon adding the charges to each system, we re-equilibrate the system volume using a Nose/Hoover barostat which leads to a slight increase in the system volume.  On average, the volume increases by 1\% for $\kappa=0.34\sigma^{-1}$ and by 3\% for $\kappa=0.086\sigma^{-1}$.  For the spheres and jacks, we also simulated larger values of $\kappa$ to explore the crossover to the limit of no salt ($\kappa \rightarrow \infty$).  It should be noted that although the DLVO potential provides an accurate representation of the forces between charged particles at large separations, the theory does not take into account the distortion of the electrical double-layer that becomes significant when a pair of particles come within two Debye lengths of each other.  We don't expect this to be a severe limitation for our dilute solutions, but it will contribute some quantitative error when comparing these results to that for real particle suspensions. 

Given the elongation of the jacks and rods, these two systems require a larger simulation cell $L$ than for the spheres and plates to avoid any issues related to the finite simulation size.  The system sizes and number of particles in each system are summarized in Table \ref{tab:size}.  Each simulation uses periodic boundary conditions in all directions.  The temperature is maintained at $T=\epsilon/k_B$ using a Langevin thermostat with damping constant $\Gamma=0.01\tau^{-1}$.  For shear simulations, the thermostat is applied only in the direction perpendicular to the bulk flow and shear directions.  
\begin{table}
 \centering
\begin{tabular}{llll}
System & $L/\sigma$ & $N_{solv}$ & $N_{part}$ \\ 
rods & 71.96 & 236367 & 143 \\ 
jacks & 72.34 & 236367 & 143 \\ 
plates & 41.34 & 44546 & 27 \\ 
spheres & 41.38 & 44550 & 27
\end{tabular} 
 \caption{Simulation box length $L$ and number of solvent particles ($N_{solv}$) and number of nanoparticles ($N_{part}$) for the four nanoparticle shapes}
 \label{tab:size}
\end{table}

Nonequilibrium molecular dynamics (NEMD) simulations are performed using the reverse perturbation method of M\"{u}ller-Plathe.\cite{FMP:PRE:99}  In this approach, particle momentum is exchanged between particles at the boundary and in the middle layer of the simulation domain to induce a shear velocity profile in the system.  The imposed momentum flux, $j_{z}(p_{x})$, and the velocity gradient, $\partial \nu_{x}/\partial z$, are recorded every 1000 simulation steps.  Using these, the viscosity for a given shear rate is calculated via
\begin{equation}
 j_{z}(p_{x})=-\eta\frac{\partial\nu_{x}}{\partial z}.
 \label{eq:momentum}
\end{equation}
      
The simulations are performed using the LAMMPS\cite{SJP:JCP:95} simulation package on quad-core 2.2 GHz AMD processors.  A total of 512 cores were used for the larger jack and rod simulations requiring about 3 hours per million simulation steps.  The sphere and plate simulations used primarily 64 cores which required about 4.5 hours per million simulation steps.  The simulations were run for up to 30 million steps or 150,000 $\tau$.

\section{Structure and Diffusion Results}
\label{sec:structure}

\begin{figure}
 \centering
 \includegraphics[clip=false,width=10 cm]{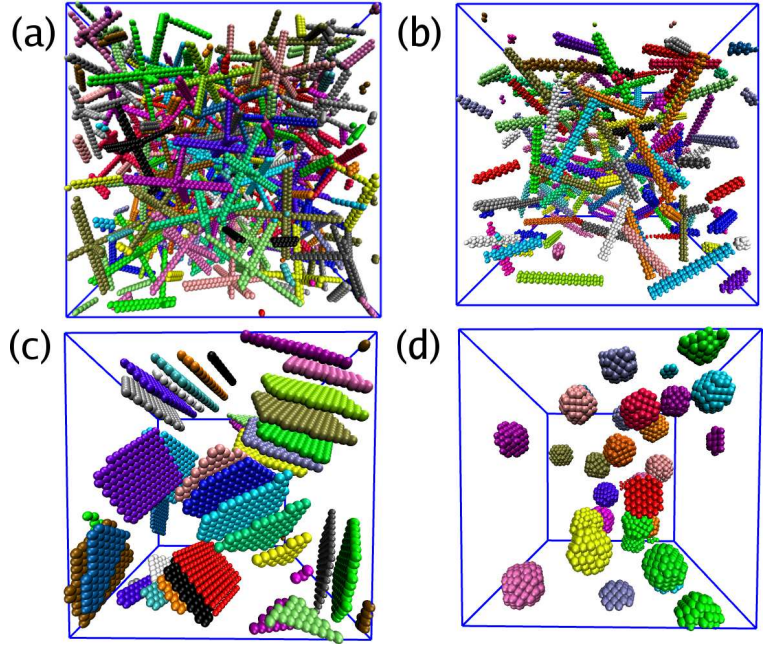}
 \caption{Snapshots of the uncharged particle suspensions for (a) jacks, (b) rods, (c) plates, and (d) spheres at $\phi_{\rm vf}=0.075$.  Solvent particles are not shown and solid particles are colored individually for visual clarity.}
 \label{fig:EQSNAPUN}
\end{figure}

\begin{figure}
 \centering
 \includegraphics[clip=false,width=10 cm]{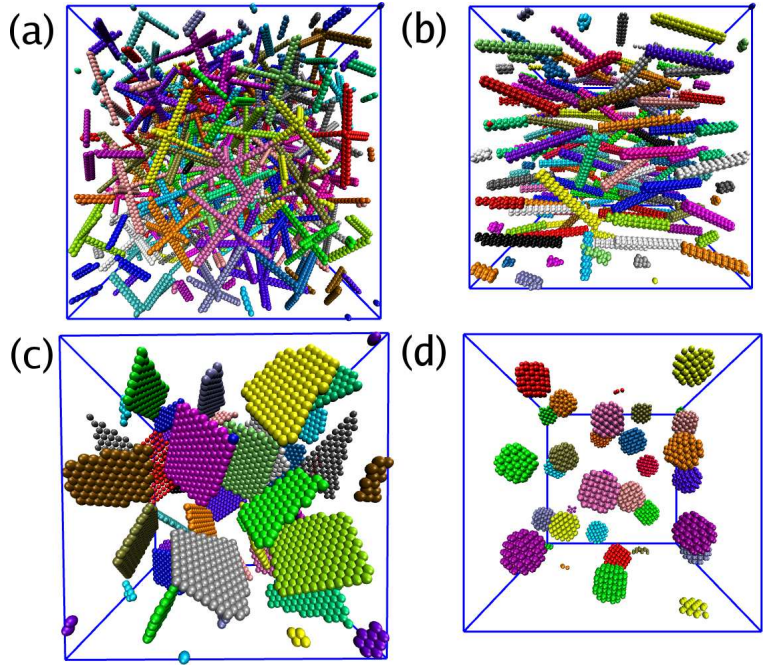}
 \caption{Snapshots of charged particle suspensions for the low salt concentration ($\kappa=0.086\sigma^{-1}$) for (a) jacks, (b) rods, (c) plates, and (d) spheres at $\phi_{\rm vf}=0.073$.  Solvent particles are not shown and solid particles are colored individually for visual clarity.}
 \label{fig:EQSNAPCH}
\end{figure}

Snapshots of equilibrated configurations of the four uncharged suspensions are shown in Fig. \ref{fig:EQSNAPUN}.  From these, we see that the jacks have a large degree of particle interactions.  The spheres have the least amount of particle interaction due to their compact geometry.  The rod and plates show a degree of particle alignment even in the absence of an external shearing force which helps to minimize the frequency of particle collisions.  Snapshots of equilibrated configurations for the charged particles in a low salt environment are shown in Fig. \ref{fig:EQSNAPCH}.  In this case the configurations of the jacks and spheres are similar to the corresponding uncharged systems whereas the rods tend to separate to maximize the distance between individual charge sites.  The plates show the greatest difference in structure.  The repulsive charge between the plates pushes apart the layered structure found in the uncharged system causing the plates to adopt a more randomly oriented structure.

\begin{figure}
 \centering
 \includegraphics[clip=false,width=8 cm]{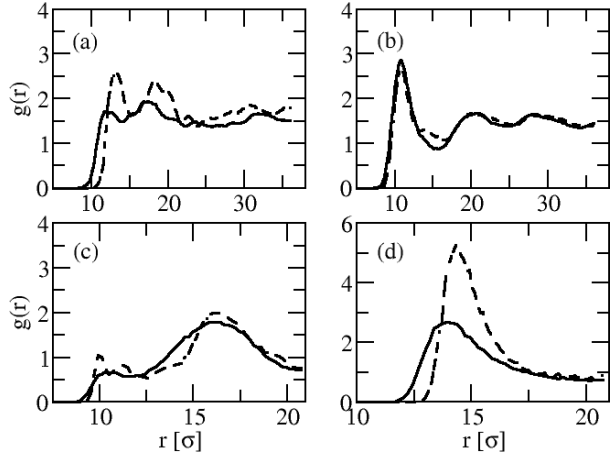}
 \caption{Radial distribution function $g(r)$ for $\kappa=0.34\sigma^{-1}$ (solid line), and $\kappa=0.086\sigma^{-1}$ (dashed line) systems for (a) jacks, (b) rods, (c) plates and (d) spheres.}
 \label{fig:RDFSUM}
\end{figure}

The change in suspension structure as a function of the shape and nanoparticle charge can be seen in the radial distribution function $g(r)$. Results for $g(r)$ measured from the center of each nanoparticle is shown in Fig. \ref{fig:RDFSUM} for the four nanoparticle shapes. As the salt concentration decreases, the amount of local order for the jacks increases.  The rods, which show a noticeable increase in orientation upon adding charge, show almost no dependence on salt concentration.  The three distinct peaks in $g(r)$ for the rods indicate a highly ordered structure for both low and high ionic strength.  The plates show slightly higher peaks when reducing the salt concentration, but not as much as the spherical particle system.  For the spherical particles, the stronger charge interaction at low ionic strength results in a larger primary peak indicating more local ordering than at high ionic strength.

\begin{figure}
 \centering
 \includegraphics[clip=false,width=8 cm]{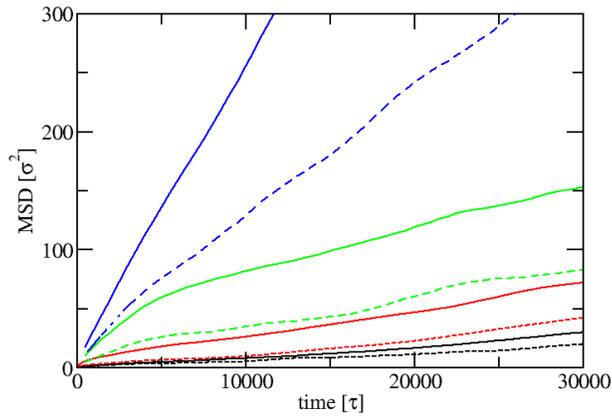}
 \caption{Mean square displacement for $\kappa=0.34\sigma^{-1}$ (solid line) and $\kappa=0.086\sigma^{-1}$ (dashed line) charged systems.  The colors correspond to jacks (black), rods (red), plates (green), and spheres (blue).}
 \label{fig:MSDSUM}
\end{figure}

\begin{figure}
 \centering
 \includegraphics[clip=false,width=8 cm]{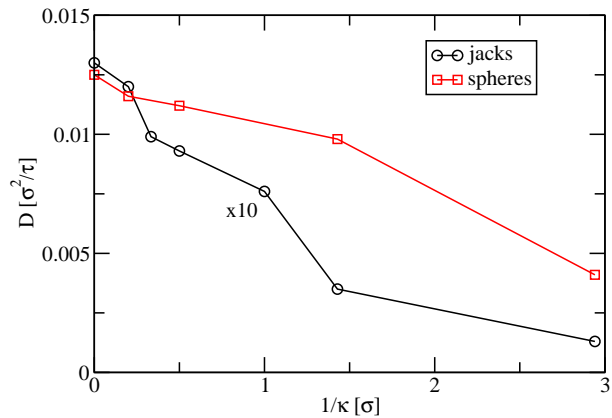}
 \caption{Diffusion constant as a function of the inverse Debye screening length $\kappa$ for the suspension of jacks (black circles) and spheres (red squares).  Results for the jacks are multiplied by 10 for clarity.}
 \label{fig:KAPPA}
\end{figure}

The change in nanoparticle shape has a strong effect on the diffusion constant $D$ as seen from the results listed in Table \ref{tab:MSD}.  The diffusion constant was determined from the slope of the mean square displacements which are shown in Fig. \ref{fig:MSDSUM} for the two charged systems.  Not surprisingly, the spheres show the highest diffusion constant followed by the plates, rods, and jacks.  Increasing the salt concentration results in a $50\%$ increase in $D$ for the jacks and a twofold increase in $D$ for the other three systems.  The low values of $D$ for the non-spherical nanoparticles is a consequence of their elongated structures and their inability to rotate freely without colliding into nearby nanoparticles.  This effect is increased as the salt concentration is reduced since the increased electrostatic repulsion leads to a larger effective nanoparticle size.  Although $D$ is roughly four times higher for the uncharged rod and plate systems compared to $D$ for the $\kappa=0.086\sigma^{-1}$ systems, the values of $D$ for the uncharged spheres are nearly seven times higher than that for the $\kappa=0.086\sigma^{-1}$ system.  To better characterize the dependence of particle diffusivity on the solution ionic strength, we performed a series of simulations varying $\kappa$ and extracted $D$.  The results, shown in Figure \ref{fig:KAPPA}, indicate that $D$ decreases fairly linearly as a function of the Debye length up to $\kappa^{-1}=3 \sigma$ suggesting that the interparticle force is sensitive to the salt concentration for this full range of $\kappa^{-1}$.  These curves will ultimately reach a plateau at higher $\kappa^{-1}$ as the electrostatic screening becomes weak enough for the particle charge to dominate the interparticle force.  In such a case, the accuracy of the idealized screened Coulomb interaction diminishes as we do not take into account effects associated with the distortions of the counterion clouds.  To further explore the role of particle shape and charge, we consider the orientational order of the rod and plate particles.

\begin{table}
\begin{center}
\caption{Diffusion constant $D$ in units of $\sigma^{2}/\tau$ for each of the four nanoparticle shapes.}
\label{tab:MSD}
\begin{tabular}{c c c c}
shape & $\kappa=0.086\sigma^{-1}$ & $\kappa=0.34\sigma^{-1}$ & uncharged \\ 
jack & 8.9e-5 & 0.00013 & 0.0013 \\ 
rod & 0.00019 & 0.00033 & 0.00079 \\ 
plate & 0.00038 & 0.00067 & 0.0016 \\ 
sphere & 0.0018 & 0.0041 & 0.0125
\end{tabular} 

\end{center}
\end{table}

The ordering in the rod and plate systems is characterized by the orientational order parameter, S, defined in Eq. \ref{eq:order} where $\theta$ is the angle between the principle axis of the particle and the director where the director is the preferred orientation.  The brackets denote an average over all particles $i$.
 
\begin{equation}
 S=\langle3cos^{2}\theta_{i}-1\rangle/2
 \label{eq:order}
\end{equation}

For each system, the orientational order parameter reached a steady state value within the equilibration period. 
The addition of charge results in an increase in S from $0.36\pm0.10$ for the uncharged rods to $S=0.67\pm0.01$ for $\kappa=0.34\sigma^{-1}$ and $S=0.70\pm0.01$ for $\kappa=0.086\sigma^{-1}$. Whereas the addition of charge results in a decrease in S for the plates from $0.24\pm0.12$ to $S=0.11\pm0.05$ for $\kappa=0.34\sigma^{-1}$ and $S=0.16\pm0.05$ for $\kappa=0.086\sigma^{-1}$.  The rods adopt a significantly more ordered conformation when charge is applied, but the plates do not.  When the plates are charged, the closely stacked aggregates shown in Fig \ref{fig:EQSNAPUN}c are forced apart into a more homogeneous but disordered configuration resulting in a reduction in the order parameter. 
 
\section{Rheology Results}
\label{sec:rheology}

\begin{figure}
 \centering
 \includegraphics[clip=false,width=10 cm]{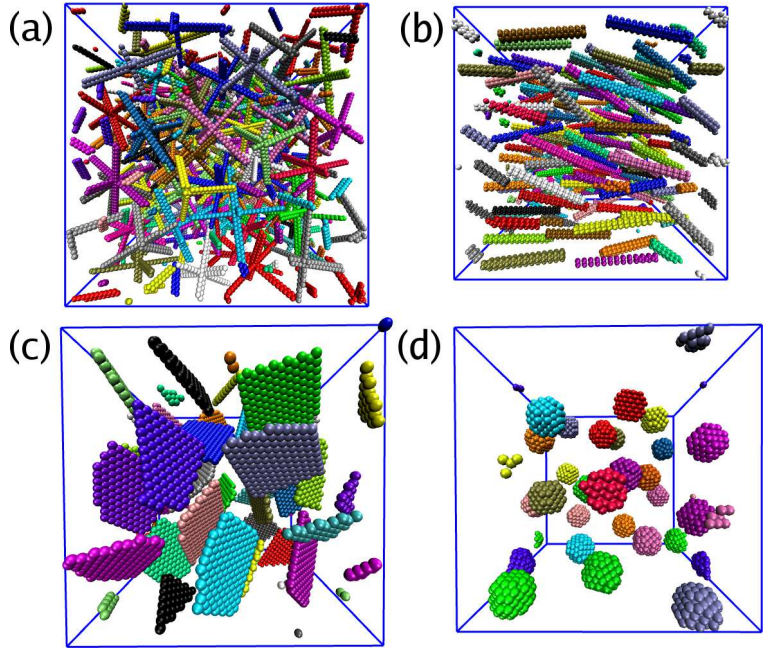}
 \caption{Simulation snapshots for the charged, low ionic strength ($\kappa=0.086\sigma^{-1}$) systems at $\phi_{\rm vf}=0.075$ under their highest simulated shear rates: (a) $\gamma=0.0046\tau^{-1}$ for jacks, (b) $\gamma=0.0020\tau^{-1}$ for rods, (c) $\gamma=0.00076\tau^{-1}$ for plates, and (d) $\gamma=0.0036\tau^{-1}$ for spheres.  In each case, the system is subject to simple shear with flow in the horizontal direction and the gradient in the vertical direction.  Solvent particles are not shown and solid particles are colored individually for visual clarity.}
 \label{fig:SHEARSNAP}
\end{figure}

Each of the fully equilibrated suspensions are sheared using the M\"{u}ller-Plathe reverse perturbation method for shear rates ranging from $7*10^{-6}$ to $0.005\ \tau^{-1}$.  Snapshots of the various systems during shear are shown in Fig. \ref{fig:SHEARSNAP} at the specified shear rates.  The particle orientations are very comparable to the unsheared systems shown in Fig. \ref{fig:EQSNAPCH} owing primarily to the fact that the particle charge causes the alignment of the rod particles and the breakup of plate aggregates even without shear.

\begin{figure}
 \centering
 \includegraphics[clip=false,width=8 cm]{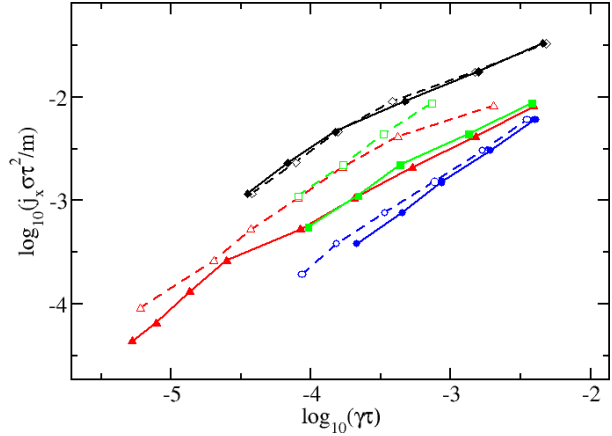}
 \caption{Log plot of momentum transferred for suspensions of various nanoparticle shapes.  The charged suspensions have a dielectric solvent with an inverse Debye screening length of $\kappa=0.34\sigma^{-1}$ (solid lines, filled symbols) or $\kappa=0.086\sigma^{-1}$ (dashed lines, open symbols) for jacks (black), rods (red), plates (green), and spheres (blue).}
 \label{fig:VISCSUM}
\end{figure}

The momentum flux as a function of shear rate for each system simulated is shown in Fig. \ref{fig:VISCSUM}.  In practice, we repeat the shear simulations for each system while systematically reducing the shear rate until the ratio of momentum flux to shear rate becomes linear.  This indicates that we have reached the zero shear viscosity for each system.  The zero shear rate viscosities are summarized in Table \ref{tab:VISC}.  The uncharged rods, and plates have nearly equal viscosities that are slightly greater than that for the spheres, similar to the observations of Knauert \textit{et al.} \cite{STK:JPS:07} from simulations at a volume fraction of $\phi_{vf}=0.05$.  For each shear rate, the viscosity of the jacks is higher than the other shapes.  Like the spheres, the jacks show little dependence on the ionic strength of the solution, but they do show a significantly higher viscosity for charged versus uncharged particles. This increase in viscosity for jacks compared to the other three shapes is most likely due to the very large effective volume occupied by the jacks, which causes them to become highly entangled as seen in Fig. 2. 
The spheres also show minimal sensitivity to the introduction of charge due to their inability to reduce particle interactions by orienting in the direction of flow. The small radius of gyration of the spheres compared to the jacks allows them to flow past each other more readily.  This results in a low viscosity for the suspensions of spheres relative to the other particle shapes.  Both the rods and plates show a much higher viscosity in the low ionic strength solvent where the charge interaction is stronger than the high ionic strength solvent.  The stronger charge interaction increases the effective diameter of the particles, which has a similar effect to increasing the volume fraction of the already constrained particles.

We also compare the viscosities of the uncharged systems to viscosities calculated using a path integral technique\cite{MLM:PRE:08} in Table \ref{tab:VISC}.  The path integral approach treats the particles as composites of spheres as they are defined in Figure \ref{fig:SINGSNAP} and determines the intrinsic viscosity by first calculating the polarizability tensor of a perfect conductor of the same shape and then relating the intrinsic viscosity to the intrinsic conductivity for an infinitely dilute solution. Assuming a linear dependence of the viscosity on the particle concentration, we find that the agreement between the continuum approach and our simulation approach ranges from very good for the spherical particles to within $50\%$ for the jacks.  This deviation for the jacks may be due to the fact that the viscosity increased non-linearly with concentration.  Since the effective volume of the jacks is large compared to the spheres, it is similar to being at a higher volume fraction such that $\eta > [\eta] c$ where $[\eta]$ is the intrinsic viscosity obtained from the path integral approach and c is the concentration.

\begin{table}
\begin{center}
\caption{Zero shear rate viscosity $\eta/\eta_{s}$, where $\eta_{s}=1.01 m/\tau\sigma$ is the viscosity of the neat solvent along with results from path integral calculations of the uncharged systems.  The value in parenthesis gives the uncertainty in the last displayed digit.}
\label{tab:VISC}
\begin{tabular}{c c c c c}
shape & $\kappa=0.086\sigma^{-1}$ & $\kappa=0.34\sigma^{-1}$ & uncharged & path integral\\ 
jack & 31(1) & 32.8(8) & 9.8(5) & 5.1(3)\\ 
rod & 14.9(6) & 8.2(5) & 1.7(2) & 2.5(1)\\ 
plate & 13.6(3) & 5.7(3) & 1.9(1) & 2.2(7)\\ 
sphere & 2.2(2) & 1.8(2) & 1.2(1) & 1.3(2)
\end{tabular} 

\end{center}
\end{table}

\begin{figure}
 \centering
 \includegraphics[clip=false,width=8 cm]{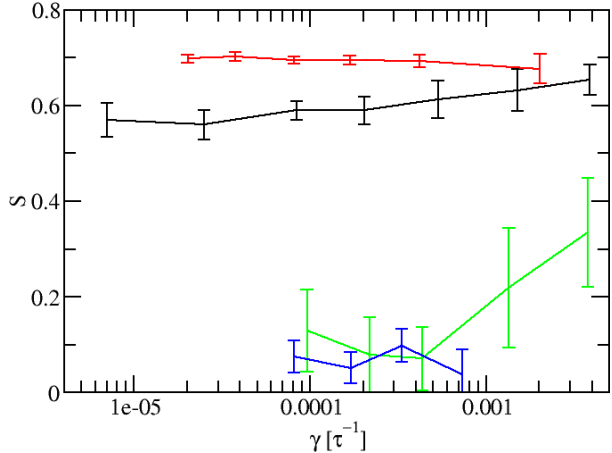}
 \caption{Orientational order parameter as a function of applied shear rate for charged rods with $\kappa=0.34\sigma^{-1}$ (black) and $\kappa=0.086\sigma^{-1}$ (red) and for charged plates with $\kappa=0.34\sigma^{-1}$ (green) and $\kappa=0.086\sigma^{-1}$ (blue).}
 \label{fig:P2SHEAR}
\end{figure}

The influence of shear flow on the orientation of these systems is shown in Figure \ref{fig:P2SHEAR} where we plot S as a function of shear rate for the charged plates and rods.  In general, shear does not induce ordering of the particle suspensions over this range of shear rates except for the highly screened plates at shear rates above $\gamma=0.001 \tau^{-1}$ and a slight increase for the highly screened rods.  Others\cite{SB:JCP:96} have shown that NEMD simulations of colloidal suspensions can produce shear induced ordering under steady shear and that the effect is dependent on the simulated system size, indicating that it is an artefact of the simulation and not a real phenomenon.  However, the shear rates shown in Figure \ref{fig:P2SHEAR} correspond to Deborah numbers ranging from $De=0.22$ to $De=174$.  In this range, artificial shear induced ordering is found to occur only at reduced temperatures below about $T^{*}=0.024$ where $T^{*}=6k_{B}T/A$.\cite{SB:JCP:96}  The reduced temperature for each of our simulations is $T^{*}=6$, so we expect that the ordering shown in Figure \ref{fig:P2SHEAR} is not an artefact of the NEMD simulation approach.

\section{Conclusions}
\label{sec:conc}

Molecular dynamics simulations were carried out to study the rheology of nanoparticle suspensions for particle shapes of jacks, rods, plates, and spheres both charged and uncharged.  We analyze the equilibrium structure of each system and find that the rod and plate systems show noticeable particle alignment which helps to minimize the frequency of particle collisions.  We expect that even more ordering in the rod and plate systems would occur at higher concentrations.  In addition, higher concentrations of jacks have exhibited enhanced viscosity\cite{MP:PRE:09} previous studies of spheres have found crystallization at volume fractions starting around $\phi=0.5$.\cite{BA:JPCM:90}  The particle diffusivity is significantly higher for the spherical particles than for all other particle shapes for both charged and uncharged particles.  We also consider the response of each particle type to imposed shear.  The viscosities of the uncharged systems are most sensitive to the ability of the particles to align under shear with the jack system having a significantly higher viscosity than the other particle shapes.  Charge effects are also most prominent for shapes that can align under shear.  Here, the presence of a repulsive charge between the particles reduces the amount of stress reduction that can be achieved by particle reorientation.  This work characterizes the role that shape plays on the rheology of charged and uncharged nanoparticle suspensions.

\section{Acknowledgements}
\label{sec:ack}

A portion of the research described in this paper was performed in the Environmental Molecular Sciences Laboratory, a national scientific user facility sponsored by the Department of Energy's Office of Biological and Environmental Research and located at Pacific Northwest National Laboratory.  This document was prepared by Corning Incorporated, as a result of the use of facilities that are operated by the U.S. Department of Energy (DOE) which are managed by Battelle.  Neither Battelle, DOE, or the U.S. Government, nor any person acting on their behalf: (a) makes any warranty or representation, expressed or implied, with respect to the information contained in this document; or (b) assumes any liabilities with respect to the use of, or damages resulting from the use of any information contained in the document.

We would like to thank the New Mexico Computing Application Center, NMCAC, for a generous allocation of computer time.  Sandia is a multiprogram laboratory operated by Sandia Corporation, a Lockheed Martin Company, for the United States Department of Energy's National Nuclear Security Administration under Contract DE-AC04-94AL85000.

\bibliographystyle{apsrev}

\end{document}